# Efficient Analysis of Pattern and Association Rule Mining Approaches

Thabet Slimani
*College of Computer Science and Information Technology , Taif University , KSA and LARODEC Lab*
*Email: thabet.slimani@gmail.com*

Amor Lazzez
*College of Computer Science and Information Technology , Taif University , KSA*
*Email: a.lazzez@gmail.com*

*Abstract*— The process of data mining produces various patterns from a given data source. The most recognized data mining tasks are the process of discovering frequent itemsets, frequent sequential patterns, frequent sequential rules and frequent association rules. Numerous efficient algorithms have been proposed to do the above processes. Frequent pattern mining has been a focused topic in data mining research with a good number of references in literature and for that reason an important progress has been made, varying from performant algorithms for frequent itemset mining in transaction databases to complex algorithms, such as sequential pattern mining, structured pattern mining, correlation mining. Association Rule mining (ARM) is one of the utmost current data mining techniques designed to group objects together from large databases aiming to extract the interesting correlation and relation among huge amount of data. In this article, we provide a brief review and analysis of the current status of frequent pattern mining and discuss some promising research directions. Additionally, this paper includes a comparative study between the performance of the described approaches.

*IndexTerms*-- Association Rule, Frequent Itemset, Sequence Mining, Pattern Mining**,** Data Mining

## 1. Introduction

Data mining [1] is a prominent tool for knowledge mining which includes several techniques: Association, Sequential Mining, Clustering and Deviation. It uses a combination of statistical analysis, machine learning and database management explore the data and to reveal the complex relationships that exists in an exhaustive manner. Additionally, Data Mining consists in the extraction of implicit knowledge (previously unknown and potentially useful), hidden in large databases.

Data mining tasks can be classified into two categories: Descriptive mining and Predictive mining. Descriptive mining refers to the method in which the essential characteristics of the data in the database are described. Clustering, Association and Sequential mining are the main tasks involved in the descriptive mining techniques tasks. Predictive mining deduces patterns from the data in a similar manner as predictions. Predictive mining techniques include tasks like Classification, Regression and Deviation detection. Mining Frequent Itemsets from transaction databases is a fundamental task for several forms of knowledge discovery such as association rules, sequential patterns, and classification [2]. An itemset is frequent if the subsets in a collection of sets of items occur frequently. Frequent itemsets is generally adopted to generate association rules. The objective of Frequent Item set Mining is the identification of items that co-occur above a user given value of frequency, in the transaction database [3]. Association rule mining [4] is one of the principal problems treated in KDD and can be defined as extracting the interesting correlation and relation among huge amount of transactions.

Formally, an association rule is an implication relation in the form X→Y between two disjunctive sets of items X and Y. A typical example of an association rule on "market basket data" is that "80% of customers who purchase bread also purchase butter ". Each rule has two quality measurements, support and confidence. The rule X→Y has confidence c if c% of transactions in the set of transactions D that contains X also contains Y. The rule has a support S in the transaction set D if S% of transactions in D contain X∪Y. The problem of mining association rules is to find all association rules that have a support and a confidence exceeding the user-specified threshold of minimum support (called MinSup) and threshold of minimum confidence (called MinConf ) respectively.

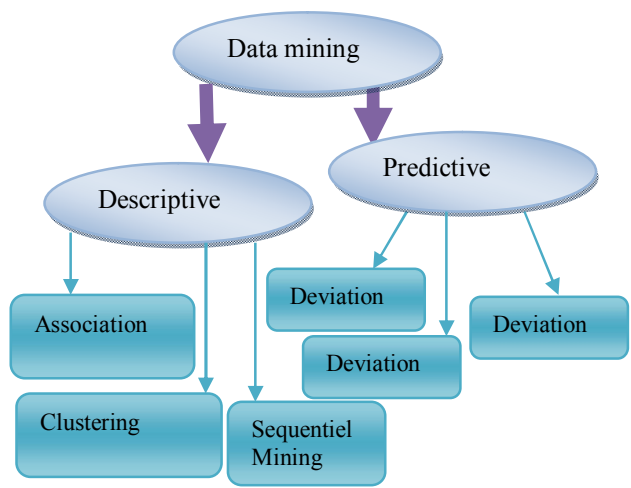

Fig 1 Data Mining tasks categories.

Actually, frequent association rule mining became a wide research area in the field of descriptive data mining, and consequently a large number of quick and speed algorithms have been developed. The more efficient are those Apriori based algorithms or Apriori variations. The works that used Apriori as a basic search strategy, they also adapted the complete set of procedures and data structures [5][6]. Additionally, the scheme of this important algorithm was also used in sequential pattern mining [7], episode mining, functional dependency discovery & other data mining fields (hierarchical association rules [8]).

This paper is organized as follows: in Section 2, we briefly describe association rules mining. Section 3 summarizes kinds of frequent pattern mining and association rule mining. Section 4 details a review of association rules approaches. In Section 5, we describe a performance analysis of the described mining algorithms. Some limited research directions are discussed in Section 6. Finally, we conclude with a summary of the paper contents.

## 2. Association Rule Mining

In this section we will introduce the association rule mining problem in detail. We will explain several concerns of Association Rule Mining (ARM).

The original purpose of association rule mining was firstly stated in [5]. The objective of the association rule mining problem was to discover interesting association or correlation relationships among a large set of data items. Support and confidence are the most known measures for the evaluation of association rule interestingness. The key elements of all *Apriori*-like algorithms is specified by the measures allowing to mine association rules which have support and confidence greater than user defined thresholds.

The formal definition of association rules is as follows: Let **I** = *i1, i2, ….im* be a set of items (binary literals). Let D be a set of database transactions where each transaction T is a set of items such that $T \subseteq I$. The identifier of each transaction is called TID. An item X is contained in the transaction T if and only if $X \subseteq T$. An association rule is defined as an implication of the form: $X \Rightarrow Y$, where $X, Y \Rightarrow T$ and $X \cap Y = \emptyset$. The rule $X \Rightarrow Y$ appears in D with support **s**, where **s** is the percentage of transactions in D that contain $X \cup Y$. The set *X* is called the antecedent and the set *Y* is called consequent of the rule. We denote by c the confidence of the rule $X \Rightarrow Y$. The rule $X \Rightarrow Y$ has a confidence **c** in D if c is the percentage of transactions in **D** containing **X** which also contain **Y**.

There are two categories used for the evaluation criteria to capture the interestingness of association rules: *descriptive* criteria (support and confidence) and *statistical* criteria. The most important disadvantage of statistical criterion is its reliance on the size of the mined population [9]. The statistical criterion requires a *probabilistic* approach to model the mined population which is quite difficult to undertake and needs advanced statistical knowledge of users. Conversely, descriptive criteria express interestingness of association rules in a more natural manner and are easy to use.

Support and confidence are the most known measures for the evaluation of association rule interestingness. In addition to the support and confidence, the quality of association rules is measured using different metric*:* the Lift criterion (LIFT) [10], the Loevinger criterion (LOEV) [11], leverage criteria [12] and Collection of quality measures is presented in [13], etc...

The support of an itemset X denoted by S(X) is the ratio of the number of transactions that contains the itemset X ($|T_X|$) to the total number of transactions ($|D|$). S(X) is defined by the following formula:

$$S(X) = \frac{|T_X|}{|D|} \quad (1)$$

The support of an association rule denoted by $S(X \Rightarrow Y)$ is the ratio of the number of transactions containing both X and Y ($|T_X \cap T_Y|$) to the total number of transactions, |D|. If the support of an association rule is 20% this means that 20% of the analyzed transactions contain $X \cup Y$. $S(X \Rightarrow Y)$ is defined by the following formula:

$$S(X \Rightarrow Y) = \frac{|T_X \cap T_Y|}{|D|} \quad (2)$$

The confidence of an association rule indicates the degree of correlation between x and y in the database. It is used as a measure of a rule's strength. The confidence of an association rule X➔Y denoted by C(X➔Y) is the ratio of the number of transactions that contain XUY (S(X➔Y)) to the number of transactions that contain X (S(X)). Consequently, if we say an association rule has a confidence of 87%, it means that 87% of the transactions containing X also contain Y. C(X➔Y) is defined by the following formula:

$$C(X \Rightarrow Y) = \frac{S(X \Rightarrow Y)}{S(X)} = \frac{|T_X \cap T_Y|}{|T_X|} \quad (3)$$

Association rule mining is described as a two-step process as follows:
- ☒ Step 1: extraction of all frequent itemsets.
- ☒ Step 2: Strong association rules extractions from the obtained frequent itemsets.

In general, association rules are considered interesting (frequent) if they satisfy both a minimum support threshold and a minimum confidence threshold defined by users or domain experts.
If the support and the confidence of an association rule X➔Y is greater than or equal to the user specified minimum support, *minsupp* and minimum confidence value, *minconf* this rule is said to be frequent (interesting). A frequent rule is characterized by the following properties:

$$S(X \Rightarrow Y) \geq minsupp \quad (4)$$

And

$$C(X \Rightarrow Y) \geq minconf \quad (5)$$

## 3. Kinds of Frequent Pattern Mining and Association Rule mining

We present in the following sections different kind of pattern to be mined and several kind of association rule mining. Several kinds of association rules mining can be defined: Frequent itemset, multilevel, multidimensional, constraints based, Boolean and quantitative association rule mining (Fig 1).

- ☒ **Frequent itemset mining**: The mining process of frequent itemsets (sets of items) can be started from transactional, relational data sets or other kinds of frequent patterns from other kinds of data sets.

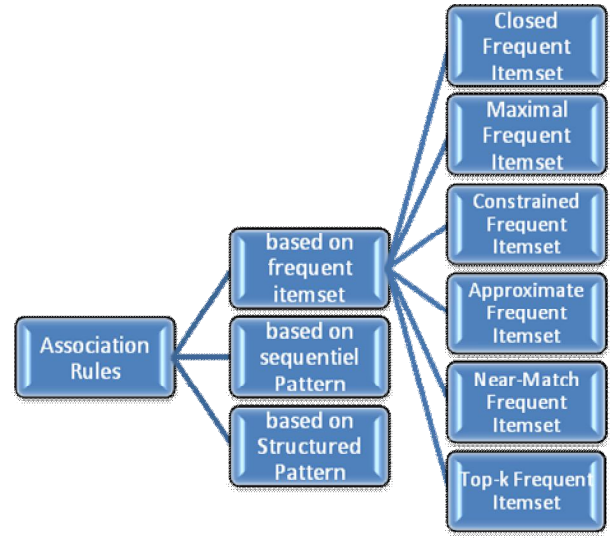

**Fig 2** Kind of pattern to be mined with Association Rules

- ☒ **Sequential pattern mining**: As an example, with sequential pattern mining, it is possible to study the order in which items are frequently purchased. Then, The mining process finds a frequent *subsequences* from a set of *sequential data set*, where a sequence records an ordering of events.
- ☒ **Structured pattern mining**: The mining process searches for frequent *substructures* in a *structured data set*. A *structure is defined* as a general concept that covers many structural forms, such as graphs, lattices, trees, sequences, sets, single items, or combinations of such structures. Consequently, structured pattern mining can be considered as the most general form of frequent pattern mining.

### 3.1 Kinds of Frequent Pattern Itemset mining

We can mine the complete set of frequent itemsets, based on the *completeness* of patterns to be mined: we can distinguish the following types of frequent itemset mining, given a minimum support threshold:

- ✓ **Closed frequent Itemset:** An itemset $X$ is a closed frequent itemset in set $S$ if $X$ is both closed and frequent in $S$.
- ✓ **Maximal frequent itemset**: An itemset $X$ is a maximal frequent itemset (or max-itemset) in set $S$ if $X$ is frequent, and there exists no super-itemset $Y$ such that $X \subset Y$ and $Y$ is frequent in $S$.
- ✓ **Constrained frequent itemset**: An itemset $X$ is a constrained frequent itemset in set S if X satisfy a set of user-defined constraints.
- ✓ **Approximate frequent itemset**: An itemset $X$ is an approximate frequent itemset in set S if X derive only approximate support counts for the mined frequent itemsets.
- ✓ **Near-match frequent itemsets**: An itemset $X$ is a near-match frequent itemset if X tally the support count of the near or almost matching itemsets.
- ✓ **Top-$k$ frequent itemset**: An itemset $X$ is a top-$k$ frequent itemset in set S if X is the $k$ most frequent itemset for a user-specified value, $k$.

### 3.2 Kinds of Association Rule Mining

Based on the *number of data dimensions* involved in the rule, we can distinguish two dimensions types of association rules:

- ☒ **Single-dimensional association rule**: An association rule is a single-dimensional, if the items or attributes in an association rule reference only one dimension. For example, if X is an itemset, then a single-dimensional rule could be rewritten as follows: *buys*(X, "*bred*")) ➔ *buys*(X, "*milk*").
- ☒ **Multidimensional association rule**: If a rule references more than one dimension, such as the dimensions *study-level, income*, and *buys*, then it is a multidimensional association rule.

Let X an itemset, the following rule is an example of a multidimensional rule:
*Study-Level*(X, "20…25")^*income*(X, "30K….40K")) ➔ *buys*(X, "*performant computer*"):

Based on the *types of values* handled in the rule, we can distinguish two types of association rules:

- ☒ **Boolean association rule**: a rule is a Boolean association rule, if it involves associations between the presence or the absence of items. For example, the following rule is a Boolean association rules obtained from market basket analysis: *buys*(X, "*computer*")) ➔ *buys*(X, "*scanner*").

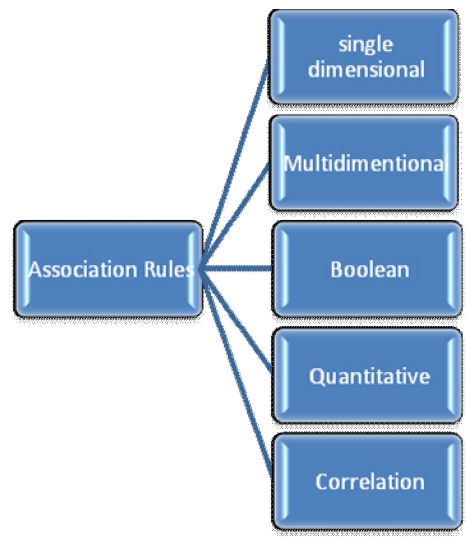

**Fig 3** Kinds of association rules mining.

- ☒ **Quantitative association rule**: a rule is a quantitative association rule, if it describes associations between quantitative items or attributes. In these rules, quantitative values for items or attributes are partitioned into intervals. For example, the following rule is a quantitative association rules:

*Study-Level*(X, "20…25")^*income*(X, "30K….40K")) ➔ *buys*(X, "*performant computer*").

Based on the *kinds of rules* to be mined, we can distinguish correlation rules defined as follows:

- ☒ **Correlation rule**: In general, such mining can generate a large number of rules, many of which are redundant or do not indicate a correlation relationship among itemsets. Consequently, the discovered associations can be further analyzed to uncover statistical correlations, leading to correlation rules.

## 4. Review of Pattern Mining Approaches

This section presents a comprehensive survey, mainly focused on the study of research methods for mining the frequent itemsets and association rules with utility considerations. Most of the existing works paid attention to performance and memory perceptions.

**Apriori:** Apriori proposed by [14] is the fundamental algorithm. It searches for frequent itemset browsing the lattice of itemsets in breadth. The database is scanned at each level of lattice. Additionally, Apriori uses a pruning technique based on the properties of the itemsets, which are: If an itemset is frequent, all its sub-sets are frequent and not need to be considered.

**AprioriTID**: AprioriTID proposed by [14]. This algorithm has the additional property that the database is not used at all for counting the support of candidate itemset after the first pass. Rather, an encoding of the candidate itemsets used in the previous pass is employed for this purpose.

**DHP:** DHP algorithm (Direct Haching and Pruning) proposed by [15] is an extension of the Apriori algorithm, which use the hashing technique with the attempts to efficiently generate large itemsets and reduces the transaction database size. Any transaction that does not contain any frequent k-itemsets cannot contain any frequent (k+1)-itemsets and such a transaction may be marked or removed.

**FDM:** FDM (Fast Distributed Mining of association rules) has been proposed by [16], which has the following distinct features.

1. The generation of candidate sets is in the same spirit of Apriori. However, some relationships between locally large sets and globally large ones are explored to generate a smaller set of candidate sets at each iteration and thus reduce the number of messages to be passed.
2. The second step uses two pruning techniques, local pruning and global pruning to prune away some candidate sets at each individual sites.
3. In order to determine whether a candidate set is large, this algorithm requires only $O(n)$ messages for support count exchange, where n is the number of sites in the network. This is much less than a straight adaptation of Apriori, which requires $O(n^2)$ messages.

**GSP**: Generalized Sequentiel Patterns (GSP) is representative Apriori-based sequential pattern mining algorithm proposed by Srikant & Agrawal in 1996 [17]. This algorithm uses the downward-closure property of sequential patterns and adopts a multiplepass, candidate generate-and-test approach.

**DIC:** This algorithm is proposed by Brin et al [18] in 1997. This algorithm partitions the database into intervals of a fixed size so as to lessen the number of traversals through the database. The aim of this algorithm is to find large itemsets which applies infrequent passes over the data than conventional algorithms, and yet uses scarcer candidate itemsets than approaches that rely on sampling. Additionally, DIC algorithm presents a new way of implication rules standardized based on both the predecessor and the successor.

**PincerSearch:** The Pincer-search algorithm [19], proposes a new approach for mining maximal frequent itemset which combines both bottom-up and top-down searches to identify frequent itemsets effectively. It classifies the data source into three classes as frequent, infrequent, and unclassified data. Bottom-up approach is the same as Apriori. Top-down search uses a new set called Maximum-Frequent-Candidate-Set (MFCS). It also uses another set called the Maximum Frequent Set (MFS) which contains all the maximal frequent itemsets identified during the process. Any itemset that is classified as infrequent in bottom-up approach is used to update MFCS. Any itemset that is classified as frequent in the top-down approach is used to reduce the number of candidates in the bottom–up approach. When the process terminates, both MFCS and MFS are equal. This algorithm involves more data source scans in the case of sparse data sources.

**CARMA:** Proposed in 1999 by Hidber [20] which presents a new Continuous Association Rule Mining Algorithm (CARMA) used to continuously produce large itemsets along with a shrinking support interval for each itemset. This algorithm allows the user to change the support threshold anytime during the first scan and always complets it at most to scan. CARMA performs Apriori and DIC on low support thresholds. Additionally CARMA readily computes large itemsets in cases which are intractable for Apriori and DIC.

**CHARM:** Proposed in 1999 Mohammed J. Zaki et al. [21] which presents an approach of Closed Association Rule Mining; (CHARM, "H" is complimentary). This effective algorithm is designed for mining all frequent closed itemsets. With the use of a dual itemset-Tidset search tree it is supposed as closed sets, and use a proficient hybrid method to skive off many search levels. CHARM significantly outpaces previous methods as proved by experimental assessment on a numerous real and duplicate databases.

**Depth-project**: DepthProject proposed by Agarwal et al., (2000) [22] also mines only maximal frequent itemsets. It performs a mixed depth-first and breadth-first traversal of the itemset lattice. In the algorithm, both subset infrequency pruning and superset frequency pruning are used. The database is represented as a bitmap. Each row in the bitmap is a bitvector corresponding to a transaction and each column corresponds to an item. The number of rows is equal to the number of transactions, and the number of columns is equal to the number of items. By using the carefully designed counting methods, the algorithm significantly reduces the cost for finding the support counts.

**FP-growth:** The principle of **FP-growth** method [23] is to found that few lately frequent pattern mining methods being effectual and scalable for mining long and short frequent patterns. FP-tree is proposed as a compact data structure that represents the data set in tree form. Each transaction is read and then mapped onto a path in the FP-tree. This is done until all transactions have been read. Different transactions that have common subsets allow the tree to remain compact because their paths overlap. The size of the FP-tree will be only a single branch of nodes. The worst case scenario occurs when every transaction has a unique itemset and so the space needed to store the tree is greater than the space used to store the original data set because the FP-tree requires additional space to store pointers between nodes and also the counters for each item.

**Eclat** : Is an algorithm proposed by Zaki [24] in 2000 for discovering **frequent itemsets** from a transaction database. The first scan of the database builds the TID_set of each single item. Starting with a single item ($k = 1$), the frequent $(k+1)$-itemsets grown from a previous $k$-itemset can be generated according to the Apriori property, with a depth-first computation order similar to FP-growth [23]. The computation is done by intersection of the TID_sets of the frequent $k$-itemsets to compute the TID_sets of the corresponding $(k+1)$-itemsets. This process repeats, until no frequent itemsets or no candidate itemsets can be found.

**SPADE**: SPADE is an algorithm for mining frequent sequential patterns **from a** sequence database proposed in 2001 by Zaki **[25].** The author uses combinatorial properties to decompose the original problem into smaller sub-problems, that can be independently solved in main-memory using efficient lattice search techniques, and using simple join operations**.** All sequences are discovered in only three database scans.

**SPAM:** SPAM is an algorithm developed by Ayres et al. in 2002 [26] for mining sequential patterns. The developed algorithm is especially efficient when the sequential patterns in the database are very long. The authors introduce a novel *depth-first* search strategy that integrates a depth-first traversal of the search space with effective pruning mechanisms. The implementation of the search strategy combines a vertical bitmap representation of the database with efficient support counting.

**Diffset** : Proposed by Mohammed J. Zaki et al. [27] in 2003 as a new vertical data depiction which keep up trace of differences in the tids of a candidate pattern from its generating frequent patterns. This work proves that diffsets is significantly expurgated (by orders of magnitude) the extent of memory needed to keep intermediate results.

**DSM-FI:** Data Stream Mining for Frequent Itemsets is a novel single-pass algorithm implemented in 2004 by Hua-Fu Li, et al. [28]. The aim of this algorithm is to excavate all frequent itemsets over the history of data streams.

**PRICES**: a skilled algorithm developed by Chuan Wang [29] in 2004, which first recognizes all large itemsets used to construct association rules. This algorithm decreased the time of large itemset generation by scanning the database just once and by logical operations in the process. For this reason it is capable and efficient and is ten times as quick as Apriori in some cases.

**PrefixSpan:** PrefixSpan proposed by Pei et al. [30] in 2004 is an approach that project recursively a sequence database into a set of smaller projected databases, and sequential patterns are grown in each projected

Efficient Analysis of Pattern and Association Rule Mining Approachesdatabase by exploring only locally frequent fragments. The authors guided a comparative study that shows PrefixSpan, in most cases, outperforms the a priori-based algorithm GSP, FreeSpan, and SPADE.

**Sporadic Rules**: Is an algorithm for mining perfectly sporadic association rules proposed by Koh & Rountreel. [31]. The authors define sporadic rules as those with low support but high confidence. They used Apriori-Inverse" as a method of discovering sporadic rules by ignoring all candidate itemsets above a maximum support threshold.

**IGB**: Is an algorithm for mining the IGB informative and generic basis of association rules from a transaction database. This algorithm is proposed by Gasmi et al. [32] in 2005. The proposal consists in reconciling between the compactness and the information lossless of the generic basis presented to the user. For this reason, the proposed approach presents a new informative generic basis and a complete axiomatic system allowing the derivation of all the association rules and a new categorization of "factual" and "implicative" rules in order to improve quality of exploitation of the knowledge presented to the user.

**GenMax:** GenMax proposed by Gouda and Zaki [33] in 2005 is a backtrack search based algorithm for mining maximal frequent itemsets. GenMax uses a number of optimizations to prune the search space. It uses a novel technique called progressive focusing to perform maximality checking, and diffset propagation to perform fast frequency computation.

**FPMax**: FPMax (Frequent Maximal Item Set) is an algorithm proposed by Grahne and Zhu, (2005) [34] based on FP Tree. It receives a set of transactional data items from relational data model, two interesting measures Min Support, Min Confidence and then generates Frequent Item Sets with the help of FPTree. During the process of generating Frequent Item Sets, it uses array based structure than tree structure. Additionally, the FPMax is a variation of the FP-growth method, for mining maximal frequent item sets. Since FPMax is a depth-first algorithm, a frequent item set can be a subset only of an already discovered MFI.

**FHARM:** Fuzzy Healthy Association Rule Mining Algorithm is a proficient algorithm developed by M. Sulaiman Khan et al. [35] in 2006. In this approach, edible attributes are filtered from transactional input data by rejections and are then converted to Required Daily Allowance (RDA) numeric values. The averaged RDA database is then converted to a fuzzy database that contains normalized fuzzy attributes comprising different fuzzy sets.

**H-Mine**: H-Mine is an algorithm for discovering frequent itemsets from a transaction database developed by Pei et al. [36] in 2007. They proposed a simple and novel data structure using hyper-links, H-struct, and a new mining algorithm, H-mine, which takes advantage of this data structure and dynamically adjusts links in the mining process. A distinct feature of the proposed method is that it has a very limited and precisely predictable main memory cost and runs very quickly in memory-based settings. Moreover, it can be scaled up to very large databases using database partitioning.

**FHSAR**: FHSAR is an algorithm for hiding sensitive association rules proposed by Weng et al. **[37].** The algorithm can completely hide any given SAR by scanning database only once, which significantly reduces the execution time. The conducted results show that FHSAR outperforms previous works in terms of execution time required and side effects generated in most cases.

**Reverse Apriori**: Is a novel algorithm presented in 2008 by Kamrul et al. [38] used in association rules mining for frequent pattern production. The proposed approach generates large frequent itemsets only if it satisfies user specified minimum item support. It then gradually decreases the number of items in the itemsets until it gets the largest frequent itemsets.

**DTFIM** : Distributed Trie-based Frequent Itemset Mining is an approach presented in 2008 by Ansari et al. [39] This algorithm is proposed for a multi-computer environment and it is revised with some FDM algorithm ideas for candidate generation step. The proposed algorithm shows that Trie data structure can be used for distributed association rule mining not just for sequential algorithms.

**GIT-tree**: GIT-tree is a tree structure developed in 2009 by [40] to mine frequent itemsets in a hierarchical database with the aim to reduce the mining time. They developed an algorithm scans database one time only and use Tidset to compute the support of generalized itemset faster.

**Scaling Apriori**: Enhanced scaling Apriori for association rule mining efficacy is developed in 2010 by Prakash & Parvathi [41] . This approach proposes an improved Apriori algorithm to minimize the number of candidate sets while generating association rules by evaluating quantitative information associated with each item that occurs in a transaction, which was usually, discarded as traditional association rules focus just on qualitative correlations. The proposed approach reduces not only the number of itemsets generated but also the overall execution time of the algorithm.

**CMRules**: CMRules is an algorithm for mining sequential rules **from** a sequence database proposed by Fournier-Viger et al. [42] in 2010. The proposed algorithm proceeds by first finding association rules to prune the search space for items that occur jointly in many sequences. Then it eliminates association rules that do not meet the minimum confidence and support thresholds according to the time ordering. The tested results show that for some datasets CMRules is faster and has a better scalability for low support thresholds.

**TopSeqRules**: TopSeqRules is an algorithm for mining sequential rules from a sequence database proposed by Fournier-Viger et al. [43] in 2010. The proposed algorithm allows to mine the top-k sequential rules from sequence databases, where $k$ is the number of sequential rules to be found and is set by the user. This algorithm is proposed, because current algorithms can become very slow and generate an extremely large amount of results or generate too few results, omitting valuable information.

**Approach based on minimum effort**: The work proposed by Rajalakshmi et al. (2011) [44] describes a novel method to generate the maximal frequent itemsets with minimum effort. Instead of generating candidates for determining maximal frequent itemsets as done in other methods [45], this method uses the concept of partitioning the data source into segments and then mining the segments for maximal frequent itemsets. Additionally, it reduces the number of scans over the transactional data source to only two. Moreover, the time spent for candidate generation is eliminated. This algorithm involves the following steps to determine the MFS from a data source:
1. Segmentation of the transactional data source.
2. Prioritization of the segments
3. Mining of segments

**FPG ARM:** Frequent Pattern Growth Association Rule Mining is an approach proposed In 2012 by Rao & Gupta [46] as a novel scheme for extracting association rules thinking to the number of database scans, memory consumption, the time and the interestingness of the rules. They used a FIS data extracting association algorithm to remove the disadvantages of APRIORI algorithm which is efficient in terms of the number of database scan and time.

**TNR:** Is an approximate algorithm developed by Fournier-Viger & S.Tseng [47] in 2012 which aims to mine the top-$k$ non redundant association rules that we name TNR (*Top-k Nonredundant R*ules). It is based on a recently proposed approach for generating association rules that is named "rule expansions", and adds strategies to avoid generating redundant rules. An evaluation of the algorithm with datasets commonly used in the literature shows that TNR has excellent performance and scalability.

**ClaSP**: ClaSP is an algorithm for mining frequent closed sequence proposed by Gomariz et al. [48] in 2013. This algorithm uses several efficient search space pruning methods together with a vertical database layout.

## 5. Performance Analysis

This section presents a comparative study, mainly focused on the study of research methods for mining the frequent itemsets, mining association rules, mining sequential rules and mining sequential pattern. Most of the existing works paid attention to performance and memory perceptions. Table 1 presents a classification of all the described approaches and algorithms.

### 5.1 frequent itemset mining

Apriori algorithm is among the original proposed structure which deals with association rule problems. In conjunction with Apriori, the AprioriTid and AprioriHybrid algorithms have been proposed. For smaller problem sizes, the AprioriTid algorithm is executed equivalently well as Apriori, but the performance degraded two times slower when applied to large problems. The support counting method included in the Apriori algorithm has involved voluminous research due to the performance of the algorithm. The proposed DHP optimization algorithm (Direct Hashing and Pruning) intended towards restricting the number of candidate itemstes, shortly following the Apriori algorithms mentioned above. The proposal of DIC algorithm is intended for database partitions into intervals of a fixed size with the aim to reduce the number of traversals through the database. Another algorithm called the CARMA algorithm (Continuous Association Rule Mining Algorithm) employs an identical technique in order to restrict the interval size to 1.

Approaches under this banner can be classified into two classes: Mining frequent itemsets without candidate generation and Mining frequent itemsets using vertical data format.

- ☒ **Mining frequent itemsets without candidate generation:** Based on the Apriori principles, Apriori algorithm considerably reduces the size of candidate sets. Nevertheless, it presents two drawbacks: (1) a huge number of candidate sets production, and (2) recurrent scan of the database and candidates check by pattern matching. As a solution, FP-growth method has been proposed to mine the complete set of frequent itemsets without candidate generation. The FP-growth algorithm search for shorter frequent pattern recursively and then concatenating the suffix rather than long frequent patterns search. Based on performance study, the method substantially reduces search time. There are many alternatives and extensions to the FP-growth approach, including H-Mine which explores a hyper-structure mining of frequent patterns; building alternative trees;

- ☒ **Mining frequent itemsets using vertical data format:** A set of transactions is presented in horizontal data format (TID, itemset), if TID is a transaction-id and itemset is the set of items bought in transaction TID. Apriori and FP-growth methods mine frequent patterns from horizontal data format. As an alternative, mining can also be performed with data presented in vertical data format. The proposed Equivalence CLASS Transformation (Eclat) algorithm explores vertical data format. Another related work with impressive results have been achieved using highly specialized and clever data structures which mines the frequent itemsets with the vertical data format is proposed by Holsheimer et al. In 1995 [49]. Using this approach, one could also explore the potential of solving data mining problems using the general purpose database management systems (dbms). Additionally, as mentioned above, the ClaSP uses vertical data format.

### 5.2 Sequential pattern mining

A sequence database contains an ordered elements or events, recorded with or without a concrete notion of time. Sequence data are involved in several applications, such as customer shopping sequences, biological sequences and Web clickstreams. As an example of sequence mining, a customer could be making several subsequent purchases, e.g., buying a PC and some Software and Antivirus tools, followed by buying a digital camera and a memory card, and finally buying a printer and some office papers. The proposed GSP algorithm includes time constraints, a sliding time window and user-defined taxonomies. An additional vertical format-based sequential pattern mining method called SPADE have been developed as an extension of vertical format-based frequent itemset mining method, like Eclat and CHARM. SPADE and GSP search methodology is breadth-first search and Apriori pruning. Both algorithms have to generate large sets of candidates in order to produce longer sequences. Another pattern-growth approach to sequential pattern mining, was PrefixSpan which works in a divide-and-conquer way. With the use of PrefixSpan, the subsets of sequential patterns mining, corresponding projected databases are constructed and mined recursively. GSP, SPADE, and PrefixSpan have been compared in [30]. The result of the performance comparison shows that PrefixSpan has the best overall performance. SPADE, although weaker than PrefixSpan in most cases, outperforms GSP. Additionally, the comparison also found that all three algorithms run slowly, when there is a large number of frequent subsequences. The use of closed sequential pattern mining can solve partially this problem.

### 5.3 Structured pattern mining

Frequent itemsets and sequential patterns are important, but some complicated scientific and commercial applications need patterns that are more complicated. As an example of sophisticated patterns we can specify : trees, lattices, and graphs. Graphs have become more and more important in modeling sophisticated structures used in several applications including Bioinformatics, chemical informatics, video indexing, computer vision, text retrieval, and Web analysis. *Frequent substructures* are the very basic patterns involving the various kinds of graph patterns. Several frequent substructure mining methods have been developed in recent works. A survey on graph-based data mining have been conducted by Washio & Motoda [50] in 2003. SUBDUE is an approximate substructure pattern discovery based on minimum description length and background knowledge was proposed by Holder et al. [51] in 1994. In addition to these studies, we list two other approaches to the frequent substructure mining problem: an Apriori-based approach and a pattern-growth approach.

## 6. Research Directions proposal

The described research under the banner of frequent pattern mining have given a solution of the most known problems related to frequent pattern mining, and the provided solutions are very good for most of the data mining tasks. But, it is required to solve several critical research problems before frequent pattern mining can become a central approach in data mining applications.

For the most current pattern mining methods, the derived set of frequent patterns is excessively massive for valuable usage. There are several propositions to reduce the huge set of patterns, which include: closed patterns, maximal patterns, condensed pattern, approximate patterns, representative patterns, clustered patterns, and discriminative frequent patterns. Additionally, much research needs to enhance the quality of preserved pattern, even it is still not clear what kind of patterns will produce the sets of pleasing pattern in both compactness and representative quality for a given application. We consider that approximate frequent patterns could be the best choice in various applications. More particularly, a mechanism of semantic frequent pattern mining approach
 (Semantic annotation for frequent patterns, and contextual analysis of frequent patterns) including a deeper understanding and interpretation of patterns is required. The semantics of a frequent pattern include deeper information : the meaning of the pattern; the patterns synonym; and the typical transactions where this pattern resides. To know the reason behind why a pattern is frequent is the main core of contextual analysis of frequent patterns over structural information. Only the work presented by Mei et al. [52] is related to the contextual analysis.

To make an improvement, it is important to analyze different properties ans solutions of the works interested by pattern mining algorithms. Based on the small subset of applications presented this article, we conclude that frequent pattern mining has claimed a broad field of applications and demonstrated its strength in solving a number of problems. We need much work to explore new applications of frequent pattern mining.

## 7. Conclusion

The most important tasks of frequent pattern mining approaches are : itemset mining, sequential pattern mining, sequential rule mining and association rule mining. A good number of efficient data mining algorithms exist in the literature for mining frequent patterns. In this paper, we have presented a brief overview of the current status and future directions of frequent pattern mining. Additionally, we have performed a comprehensive study of some algorithms and methods that exists for the mining of frequent patterns. With over a decade of extensive research, a good number of research publications, development and application activities in this domain have been proposed. We give a brief discussion of a number of algorithms presented along this decade with a comparative study of a few significant ones based on their performance. However, we require to conduct a deep research based on several critical issues so that this domain may have its factual existence and deep impact in data mining applications.

Efficient Analysis of Pattern and Association Rule Mining Approaches

| Approach | year | Data source | Sequentiel Pattern Mining | Sequentiel Rule Mining | Frequent Itemset Mining | Association Rule Mining |
|---|---|---|---|---|---|---|
| Apriori | 1994 | Transaction database | | | √ | |
| AprioriTID | 1994 | Transaction database | | | √ | |
| DHP | 1995 | Transaction database | | | √ | |
| FDM | 1996 | Transcation database | | | √ | |
| GSP | 1996 | Sequence Database | √ | | | |
| DIC | 1997 | Transcation database | | | √ | |
| PincerSearch | 1998 | Transcation database | | | √ | |
| CARMA | 1999 | Transcation database | | | √ | |
| CHARM | 1999 | Transcation database | | | √ (closed) | |
| Depth-project | 2000 | Transcation database | | | √ (maximal) | |
| Eclat | 2000 | Transcation database | | | √ | |
| SPAD | 2001 | Sequence Database | √ | | | |
| SPAM | 2002 | Sequence Database | √ | | | |
| Diffset | 2003 | Transcation database | | | √ | |
| FP-growth | 2004 | Transcation database | | | √ | FP-growth |
| DSM-FI | 2004 | Transcation database | | | √ | |
| PRICES | 2004 | Transaction database | | | √ | |
| PrefixSpan | 2004 | Sequence Database | √ | | | |
| Sporadic Rules | 2005 | Transaction database | | | | √ |
| IGB | 2005 | Transaction database | | | | √ |
| GenMax | 2005 | Transaction database | | | √ (maximal) | |
| FPMax | 2005 | Transaction database | | | √ (maximal) | |
| FHARM | 2006 | Transaction database | | | √ | |
| H-mine | 2007 | Transaction database | | | √ | |
| FHSAR | 2008 | Transaction database | | | | √ |
| Reverse Apriori | 2008 | Transaction database | | | √ (maximal) | |
| DTFIM | 2008 | Transaction database | | | √ | |
| GIT-tree | 2009 | Transaction database | | | √ | |
| Scaling Apriori | 2010 | Transaction database | | | √ | |
| CMRules | 2010 | Sequentiel database | | √ | | |
| Minimum effort | 2011 | Transaction database | | | √ (maximal) | |
| TopSeqRules | 2011 | Sequentiel database | | √ | | |
| FPG ARM | 2012 | Transaction database | | | √ | |
| TNR | 2012 | Transaction database | | | | √ |
| ClaSP | 2013 | Sequentiel database | √ (closed) | | | |

## Authors' Profiles

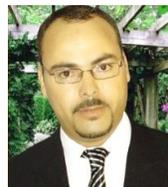

**Thabet Slimani:** got a PhD in Computer Science (2011) from the University of Tunisia. He is currently an Assistant Professor of Information Technology at the Department of Information Technology of Taif

University at Saudia Arabia and a LARODEC Labo member (University of Tunisia), where he is involved both in research and teaching activities. His research interests are mainly related to Semantic Web, Data Mining, Business Intelligence, Knowledge Management and recently Web services. Thabet has published his research through international conferences, chapter in books and peer reviewed journals. He also serves as a reviewer for some conferences and journals.

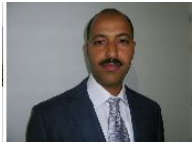

**Amor Lazzez:**

is currently an Assistant Professor of Computer an Information Science at Taif University, Kingdom of Saudi Arabia. He received the Engineering diploma with honors from the high school of computer sciences (ENSI), Tunisia, in June 1998, the Master degree in Telecommunication from the high school of communication (Sup'Com), Tunisia, in November 2002, and the Ph.D. degree in information and communications technologies form the high school of communication, Tunisia, in November 2007. Dr. Lazzez is a researcher at the Digital Security research unit, Sup'Com. His primary areas of research include design and analysis of architectures and protocols for optical networks.